\documentclass[aps,pre,reprint,superscriptaddress,floatfix,longbibliography]{revtex4-2}

\usepackage[T1]{fontenc}
\usepackage[utf8]{inputenc}
\usepackage[american]{babel}
\usepackage{amsmath}
\usepackage{amssymb}
\usepackage{graphicx}
\usepackage{hyperref}
\usepackage{color}

\begin{document}

\title{The cubic discriminant as an organizing principle for Duffing dynamics}

\author{Patrycja Jaros}
\author{Przemyslaw Perlikowski}
\affiliation{Division of Dynamics, Lodz University of Technology,
Stefanowskiego 1/15, 90-924 Lodz, Poland}

\date{\today}

\begin{abstract}
We present an analytical organizing principle underlying the dynamics of
the periodically forced Duffing oscillator. In the massless limit the
system reduces to a time-dependent cubic equation whose discriminant and
Jacobian classify the canonical Duffing regimes and determine the number
and stability of instantaneous equilibria. For the double-well
oscillator, the same construction gives an exact forcing threshold
$F_{\mathrm{bif}}$ separating intra-well and inter-well dynamics. We
show that this algebraic quantity governs the finite-mass dynamics far
beyond the singular limit from which it originates: the boundary between
intra-well and inter-well motion converges to $F_{\mathrm{bif}}$ as
$m\to0$, and the families of pitchfork bifurcations responsible for
asymmetric inter-well states accumulate at a common point. Numerical
simulations further show how the discontinuous jumps of the degenerate
model are regularized at finite mass into short oscillatory transients
that shrink as $m\to0$. These results reveal the massless Duffing
equation as the organizing center of the full finite-mass dynamics and
establish a direct link between the algebraic structure of the
degenerate problem and the global bifurcation structure of the
oscillator.
\end{abstract}

\maketitle

\section{Introduction}

The Duffing oscillator is one of the canonical models of nonlinear
physics and remains a standard reference system for the study of forced
and damped oscillations with strong
nonlinearity~\cite{holmes1979nonlinear,kovacic2011duffing}, capturing
phenomena that range from vibrations of buckled
structures~\cite{moon1979magnetoelastic} to nonlinear electronic
circuits~\cite{ueda1979randomly}, nanomechanical
resonators~\cite{aldridge2005noise} and bistable devices for energy
harvesting~\cite{cottone2009nonlinear,erturk2009piezomagnetoelastic},
and providing the standard setting for stochastic
resonance~\cite{gammaitoni1998stochastic}. Despite its structural
simplicity, it exhibits rich dynamical behavior, including
multistability, symmetry breaking, period-doubling cascades and
deterministic
chaos~\cite{ueda1979randomly,holmes1979nonlinear,parlitz1993common}.
Its bifurcation set possesses a remarkably regular superstructure,
shared by a wide class of driven dissipative oscillators and classified
in detail with the help of torsion and winding numbers of periodic
orbits~\cite{parlitz1985superstructure,parlitz1986resonances,parlitz1987period,scheffczyk1991comparison,englisch1991regular,englisch1994winding,medeiros2013torsion,englisch2015comparison}.

Among the classical Duffing variants, the double-well oscillator is of
particular interest because it combines local oscillations within a
single potential well with global motion connecting both wells.
Depending on forcing and damping, trajectories may remain confined to
one well (intra-well motion) or repeatedly cross the potential barrier
(inter-well motion), a distinction that is fundamental for transport,
escape phenomena, basin structure and multistability in bistable
systems~\cite{moon1985fractal,thompson1989chaotic,mcdonald1985fractal,kim2000dynamic,gilmore1995structure}.
Analytical results for this system concern mainly the onset of chaotic
transients and the erosion of safe basins, obtained through
Melnikov-type homoclinic
criteria~\cite{holmes1979nonlinear,guckenheimer1983nonlinear,moon1985fractal,thompson1989chaotic}.
The boundary between intra-well and inter-well motion itself is,
however, almost invariably located by numerical bifurcation analysis,
which identifies where the transition occurs but not why particular
parameter values play a special role. The existence of a simple analytical characterization of this boundary remains largely unexplored. 

In this work we show that such an origin exists and resides in the
massless limit ($m\to0$) of the Duffing equation. When the inertia tends to zero,
the differential equation degenerates into a time-dependent cubic
algebraic equation, whose discriminant determines the number of
instantaneous equilibria and changes sign at a precisely defined forcing
amplitude. We demonstrate that this apparently singular
limit is not merely a mathematical curiosity
but acts as an organizing center for the full finite-mass dynamics and has direct dynamical consequences.

The central result of the paper is that the vanishing of the cubic
discriminant defines an exact forcing amplitude threshold $F_{bif}$,
which separates qualitatively different dynamical regimes in the
double-well Duffing oscillator. We show that the finite-mass boundary
between intra-well and inter-well motion converges to this value as
$m\to0$ and that the families of pitchfork bifurcations responsible for
the creation of asymmetric inter-well states accumulate onto the single
point $(m,F)=(0,F_{bif})$, revealing a direct connection between the
algebraic structure of the degenerate problem and the global bifurcation
structure of the full system.

\begin{figure*}[t]
\includegraphics[width=0.65\textwidth]{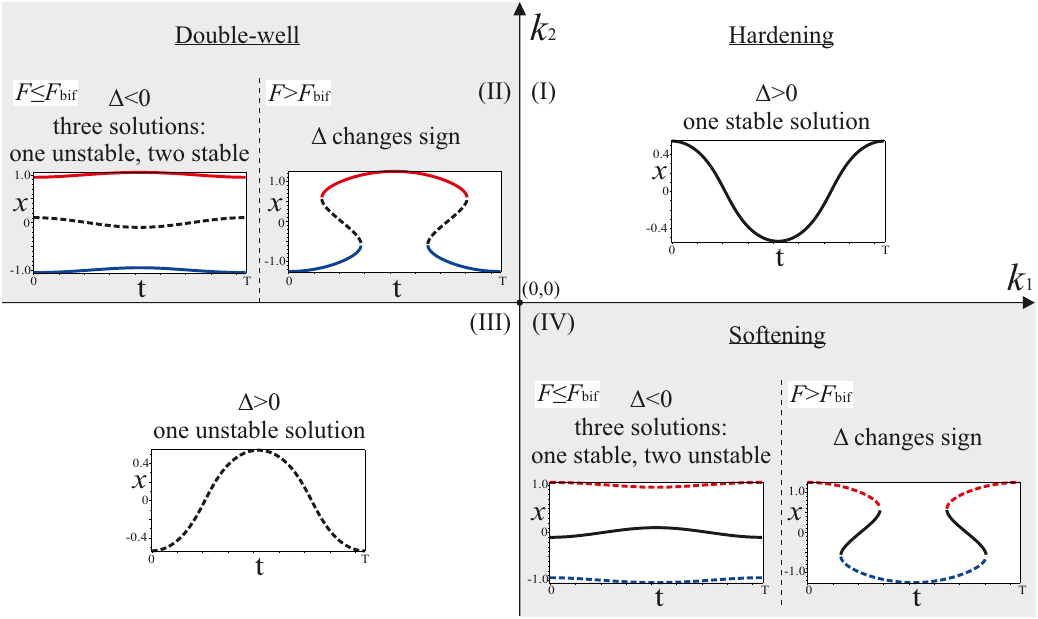}
\caption{Quadrant number: (I) hardening Duffing with $k_{1}=1$ and $k_{2}=1$; (II)
double-well Duffing with $k_{1}=-1$ and $k_{2}=1$; (III) $k_{1}=-1$ and $k_{2}=-1$; and
(IV) softening Duffing with $k_{1}=1$ and $k_{2}=-1$. The insets show exemplary time
plots (calculated as roots of the polynomial Eq.~(\ref{eq:deg})) for the degenerate
massless case. The behavior of the system in each case is strictly related to the sign
of the discriminant $\Delta$.}\label{fig:cases}
\end{figure*}

Finally, we show numerically how the discontinuous switching predicted
by the massless model emerges at finite mass: after each jump the full
system performs decaying oscillations whose duration shrinks with $m$.
The cubic discriminant is thus not merely a property of the degenerate
equation but a quantity that leaves a measurable footprint in the
dynamics for finite inertia, providing an unexpected organizing
principle for Duffing dynamics.

\section{THE MASSLESS LIMIT}\label{sec:Reduction-to-the}
\subsection{Discriminant classification}
The system under study is the Duffing oscillator,
\begin{equation}
m\ddot{x}+c\dot{x}+k_{1}x+k_{2}x^{3}=F\cos\left(\omega t\right),\label{eq:main}
\end{equation}
with mass $m$, linear stiffness $k_{1}$, cubic stiffness $k_{2}$, forcing amplitude $F$
and forcing frequency $\omega$. The damping coefficient is set to $c=0.2\sqrt{m}$, which
keeps the relative damping ratio independent of mass,
allowing us to vary $m$ without changing the effective level of damping in the system.
We assume that the values of stiffness are constant.

Duffing behavior is strictly dependent on the choice of the linear and cubic spring
stiffnesses $k_{1}$ and $k_{2}$. Depending on the signs of those parameters the system
dynamics changes completely. This is also evident for a degenerate system with zero
mass. Hence, let us focus on the degenerate case with $m=0$. As in such a
situation $c=0.2\sqrt{m}$ also vanishes, the system transforms into a third-degree
polynomial with a time-dependent constant term $F\cos\left(\omega t\right)$,
\begin{equation}
k_{1}x+k_{2}x^{3}-F\cos\left(\omega t\right)=0.\label{eq:deg}
\end{equation}
Such a form of the cubic equation has a different number of real solutions depending on
the discriminant,
\begin{equation}
\Delta=\frac{1}{27}\left(\frac{k_{1}}{k_{2}}\right)^{3}+\frac{F^{2}\cos^{2}(\omega t)}{4k_{2}^{2}}.\label{eq:discriminant}
\end{equation}
If $\Delta>0$, Eq.~(\ref{eq:deg}) has one real and two complex solutions; if
$\Delta<0$ it has three distinct real solutions; and at $\Delta=0$ two of them
coincide while the third remains simple. (Here $\Delta$ is the radicand of Cardano's formula,
so its sign is opposite to the classical cubic discriminant: $\Delta>0$ corresponds to a
single real root.)

Exact solutions of Eq.~(\ref{eq:deg}) can be obtained with the help of Cardano's
formula, while their stability can be investigated by analyzing the sign of the Jacobian
obtained for the system
\begin{equation}
c\dot{x}+k_{1}x+k_{2}x^{3}=F\cos\left(\omega t\right).\label{eq:deg-1}
\end{equation}
The damping coefficient here is assumed to be a small value $c=\varepsilon>0$, which is
correct remembering that $c=0.2\sqrt{m}$, so as $m\rightarrow0$ then $c\rightarrow0$, but
noticeably slower. The sign of the Jacobian of Eq.~(\ref{eq:deg-1}) does not depend on
the parameter $c$, as it is given by the expression
\begin{equation}
J(x(t))=\frac{1}{c}\left(-k_{1}-3k_{2}(x(t))^{2}\right).\label{eq:jac}
\end{equation}
The sign of $k_1/k_2$ organizes the four cases sketched in Figure~1. For
  $k_1/k_2>0$ (quadrants I and III) $\Delta>0$ for all $t$, so Eq.~(2) has a
  single root; the Jacobian~(5) is negative in I (stable) and positive in III
  (unstable). For $k_1/k_2<0$ the discriminant is negative for all $t$ -- three
  real roots throughout the cycle -- provided the forcing stays below the
  critical amplitude calculated for $\Delta=0$:

\begin{equation}
 F_{bif}=\frac{2\sqrt{3}}{9}\sqrt{\dfrac{|k_{1}|^{3}}{|k_{2}|}}.\label{eq:fbif}
\end{equation}
Above $F_{bif}$ the sign of $\Delta$ fluctuates, so Eq.~(2) alternates between
  three roots and one over the cycle. The two subcases differ only through the
  Jacobian: in the double-well case (II) the middle branch is unstable and the
  two well branches stable, whereas in the softening case (IV) the middle branch
  is stable and the outer two unstable. The degenerate cubic thus already
  reproduces the qualitative behavior of the full Duffing oscillator.

\subsection{Degenerate double-well}

\begin{figure}[t]
\includegraphics[width=\columnwidth]{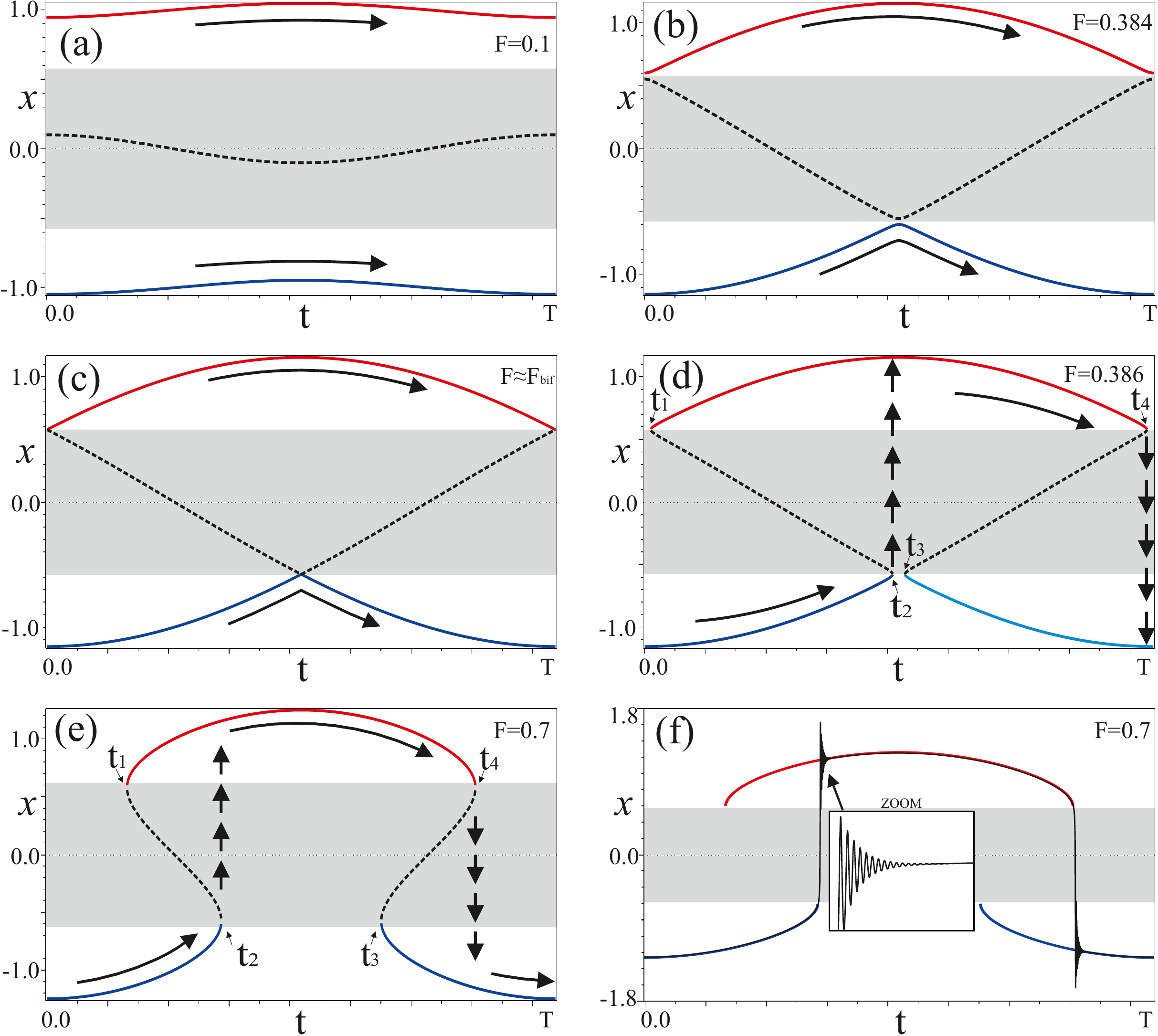}
\caption{Instantaneous root branches of Eq.~(\ref{eq:deg}) obtained analytically for
$k_{1}=-1$ and $k_{2}=1$. Red and blue curves denote stable (attracting) root branches, dashed black the unstable (repelling) one, in the sense of the small-$c$ regularization (\ref{eq:deg-1}). The gray background marks the region where
the Jacobian is positive and black arrows indicate the direction of
motion along the root branches. (a) $F=0.1$ -- each root branch has small amplitude;
(b) $F=0.384$ -- close to the bifurcation point every root branch grows in
amplitude, nearly reaching the gray region; (c)
$F=F_{bif}=\frac{2\sqrt{3}}{9}$ -- the unstable root branch touches each of the
stable root branches at a single point; (d) $F=0.386$ and (e) $F=0.7$ -- the occupied
stable root branch loses continuity and the trajectory jumps to the other
stable root branch at $t_{2}$ and $t_{4}$, while at $t_{1}$ and $t_{3}$ a
new pair of roots is born and no jump occurs; (f) comparison of the
analytical solution at $F=0.7$ with the numerical solution of the full
model Eq.~(\ref{eq:main}) for $m=10^{-3}$ (continuous black line).}\label{fig:deg}
\end{figure}

We consider in more detail the case from quadrant (II) with $k_{1}=-1$ and $k_{2}=1$
(double-well Duffing). The corresponding instantaneous roots of the degenerate Eq.~(\ref{eq:deg})
are illustrated in Figure~\ref{fig:deg}, which provides an overview of the qualitative
behavior discussed below.

For the considered case $\Delta=-\frac{1}{27}+\frac{F^{2}\cos^{2}(\omega t)}{4}$. Taking
into account the properties of the cubic function, one can demonstrate that for the
control parameter $F\leq\frac{2\sqrt{3}}{9}$ Eq.~(\ref{eq:deg}) consistently possesses
three roots [Figure~\ref{fig:deg}(a-c)]. In contrast, when a forcing term $F$
greater than $\frac{2\sqrt{3}}{9}$ is applied, the function in Eq.~(\ref{eq:deg}) has
either one or three real roots, depending on the moment in time
[Figure~\ref{fig:deg}(d,e)]. The boundary case $F_{bif}=\frac{2\sqrt{3}}{9}$ is presented in
Figure~\ref{fig:deg}(c). 

The Jacobian indicates that root branches $x(t)$ lying within the interval
$[-\frac{\sqrt{3}}{3},\frac{\sqrt{3}}{3}]$ are unstable, whereas outside this interval
they are stable. Consequently, there is one unstable root branch - the one oscillating
(with amplitude at most $\frac{\sqrt{3}}{3}$) around zero - while the two root branches
corresponding to the positive and negative potential wells are stable.

For $F>\frac{2\sqrt{3}}{9}$ it turns out that at the moment when the discriminant
$\Delta$ becomes zero, the  trajectory is thrown to the opposite root branch due to discontinuity of the root branches (transitions between the two potential wells). Writing $\varphi\equiv\arccos(F_{bif}/F)$, the discriminant vanishes four times
  per period, at $\omega t_1=\varphi,\ \omega t_2=\pi-\varphi,\ \omega t_3=\pi+\varphi$ and $\omega t_4=2\pi-\varphi$,
  yet only two produce jumps. At $t_1$ and $t_3$ a root pair is born
  at $x=\pm\sqrt3/3$, leaving the occupied root branch  undisturbed; at $t_2$ and
  $t_4$ the occupied root branch merges with the middle one and vanishes,
  throwing the trajectory into the opposite root branch [Fig.~2(d,e)].

The results presented in Figure~\ref{fig:deg} correspond  to those obtained for
the non-degenerate system~(\ref{eq:main}) with $m\rightarrow0$. For
$F\leq\frac{2\sqrt{3}}{9}$ (intra-well cases: Figures~\ref{fig:deg}(a)-(c)) the
correspondence is exact; for
$F>\frac{2\sqrt{3}}{9}$ the system displays more complex behavior and validation can
be obtained only through the limit analysis, since after the jump at $t_{2}$ or $t_{4}$
the full model~(\ref{eq:main}) with $m=10^{-3}$ performs several oscillations [see inset
in Figure~\ref{fig:deg}(f)].  To achieve timeplot integration of full model Eq.~(\ref{eq:main}) is performed with a Runge--Kutta method with adaptive step size over the interval $[0,T]$, where $T=2\pi/\omega$. Initial condition corresponds to $t=0$ of analytical solution.

\subsection{Convergence to degenerate solution}

Figure~3(a) reports the duration $\Delta t$ of post-jump oscillations for
displacement and velocity as a function of $m$. $\Delta t$ decreases with $m$
following an approximate power law ($\Delta t\propto m^{\alpha}$ with $\alpha\approx0.45$) for both displacement and velocity, and for
the smallest considered mass the solution approaches the discontinuous jump predicted by
the degenerate analysis.

\begin{figure}[b]
\includegraphics[width=\columnwidth]{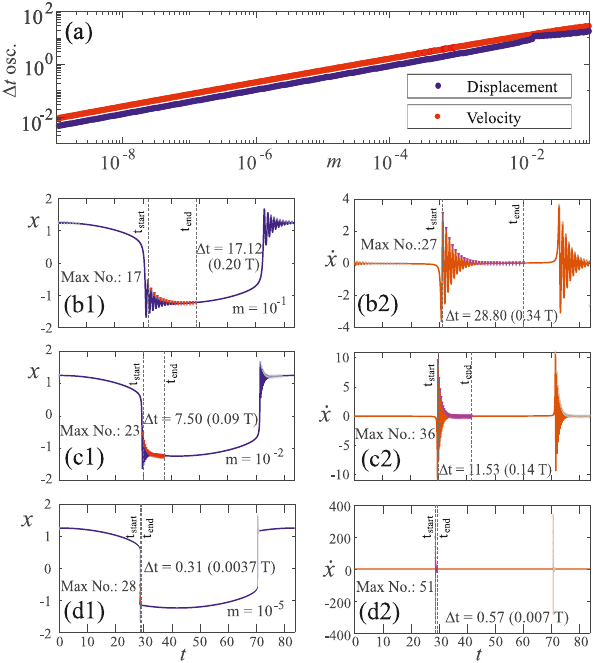}
\caption{(a) Time interval of oscillations for displacement and velocity with
respect to mass $m$. (b1-d1) time plot of displacement $x$; (b2-d2) time plot of
velocity $\dot{x}$. Chosen mass values: (b1,b2) $m=10^{-1}$, (c1,c2) $m=10^{-2}$,
(d1,d2) $m=10^{-5}$. $t_{\mathrm{start}}$ and $t_{\mathrm{end}}$ mark the first and last maximum of the
post-jump oscillation. $\Delta t = t_{\mathrm{end}}-t_{\mathrm{start}}$. $\omega=0.075$; $F=0.7$.}\label{fig:Tor1}
\end{figure}

 The procedure for determining the
oscillation duration is illustrated for $m=10^{-1}$, $m=10^{-2}$ and $m=10^{-5}$ in 
panels~(b1), (c1) and (d1) for displacement, and in panels~(b2), (c2) and (d2) for
velocity. For each trace we detect the local maxima around the occupied (left) well; for
  smaller $m$ their count may differ between displacement and velocity, since the
  velocity remains slightly positive and suppresses some displacement maxima.  In each panel we
report the value of $\Delta t$ both in time units and (in brackets) as a fraction of the
period $T$, along with the number of detected local maxima.

In panels~(d1,d2) we observe that the finite-mass trajectories converge
towards the discontinuous switching solution predicted by the degenerate
analysis. The oscillatory transients generated after each transition
between wells become progressively shorter as the inertia decreases,
demonstrating how the finite-mass dynamics approaches the massless
limit. The discontinuous jumps of the degenerate model are therefore not
an isolated feature of the singular problem but the limiting form of the
switching dynamics of the full Duffing oscillator.

\section{ORGANIZING STRUCTURE OF THE FINITE-MASS SYSTEM}

Throughout this section, as a representative of the case $k_{1}<0$, $k_{2}>0$, we choose $k_{1}=-1$, $k_{2}=1$ and $c=0.2\sqrt{m}$. Moreover, unless stated otherwise, $\omega=0.075$. All numerical integrations of Eq.~(\ref{eq:main}) are performed with a
Runge--Kutta method with adaptive step size.

\begin{figure}[t]
\includegraphics[width=\columnwidth]{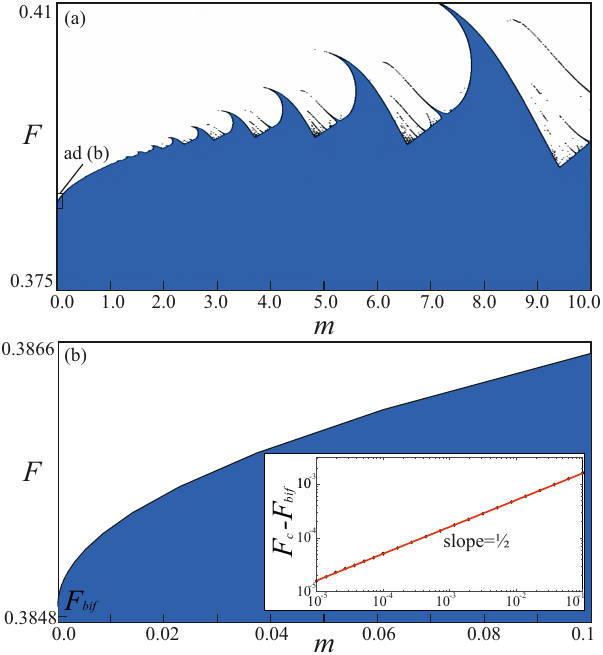}
\caption{(a) Existence of stable intra-well attractors in the $(m,F)$ parameter plane. Blue region corresponds to
parameter values for which at least one stable intra-well attractor exists, while white
regions indicate its absence. (b) zoom of region indicated in panel (a). In the inset in panel (b) log-log scale diagram with red line confirming the critical forcing amplitude
$F_{bif}=2\sqrt{3}/9$ predicted by the cubic discriminant. } \label{fig:intra}
\end{figure}

Section~\ref{sec:Reduction-to-the} identifies a single quantity, the massless threshold
$F_{bif}$ of Eq.~\eqref{eq:fbif}, that separates intra-well/inter-well regimes in the degenerate limit. It leaves two independent footprints in the finite-mass
dynamics. First, the boundary between intra-well and inter-well motion in the $(m,F)$
plane converges to $F_{bif}$ as $m\to0$; second, the pitchfork bifurcations that generate
the asymmetric inter-well solutions accumulate onto the single point $(0,F_{bif})$. Neither
signature refers to the degenerate equation, yet both single out the same forcing amplitude.

\subsection{Intra-well boundary}

First, Figure~\ref{fig:intra} shows the region of existence of stable intra-well attractors
in the $(m,F)$ parameter plane. A point is classified as intra-well when a trajectory started in the well stays
there after the transient ($x(t)$ never crosses $x=0$); $F_{c}(m)$ is the boundary
of the region where such a confined attractor exists.

The boundary between intra-well and inter-well motion converges toward $F_{bif}$ as
$m\rightarrow0$. This demonstrates that the discriminant threshold persists beyond the
degenerate limit and provides the first indication that the massless system acts as an
organizing structure of the full Duffing dynamics. In panel (a) the boundary is presented for $m\in(0,10]$, while panel (b) zooms into the vicinity of $m=0$, $m\in(0,0.1]$. The inset shows the convergence to $F_{bif}$ on a log-log scale. The slope is equal $1/2$ which means that $F_c-F_{bif}\propto\sqrt{m}$ as $m\to 0$. Here $F_c$ denotes  the critical (boundary) value of $F$, being a function of $m$. We note that the exponent $1/2$ is tied to the adopted damping
convention $c=0.2\sqrt{m}$, so the observed scaling is naturally
interpreted as $F_{c}-F_{bif}\propto c$, with the $\sqrt{m}$ law
inherited from the relation $c(m)$.

\subsection{Accumulation of pitchfork tongues}

\begin{figure}[t]
\includegraphics[width=\columnwidth]{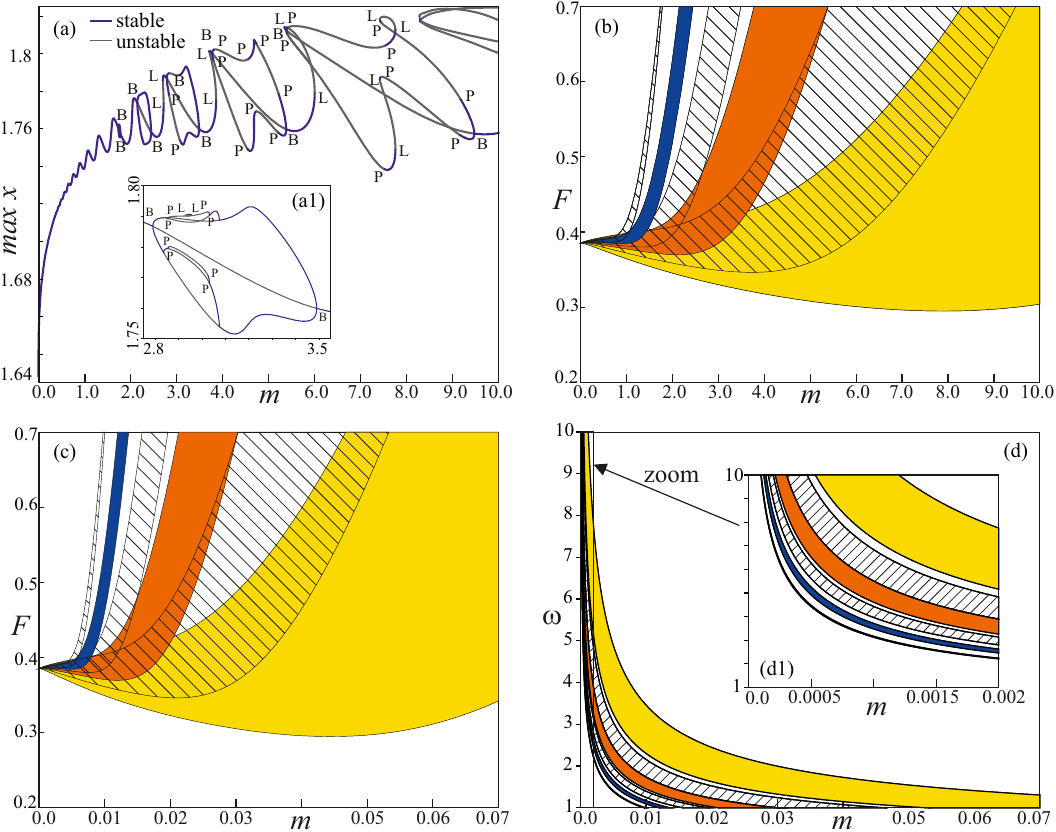}
\caption{(a) One-dimensional continuation with fixed $\omega=0.075$, $F=0.7$
and varied $m$. In (b,c) two-parameter continuations of the pitchfork tongues
in the $(m,F)$ plane are shown for $\omega=0.075$ and $\omega=1.0$,
respectively. In (d) the two-parameter continuation of those tongues is
presented in the $(m,\omega)$ space for fixed $F=0.7$. Insets (a1) and (d1)
are magnifications of panels (a) and (d). Colors and hatching serve only to
distinguish successive tongues.} \label{fig:BifAuto07} 
\end{figure}

Second, the important role of the bifurcation point
$(m,F)=(0,F_{bif})=(0,\frac{2\sqrt{3}}{9})$ identified in
Section~\ref{sec:Reduction-to-the} is  evident when examining the bifurcation
structure. 
The stable and unstable periodic solutions and their bifurcations are
computed by numerical continuation in one (mass $m$) and two (mass $m$ and amplitude $F$ or frequency $\omega$) parameters using
AUTO-07p~\cite{doedel2007auto}. In Figure~\ref{fig:BifAuto07} we show the bifurcation structure of inter-well motion  and its robustness taking not only $\omega=0.075$ (panels (a,b)), but also $\omega=1.0$ (panel (c)) and the whole range of $\omega\in[1,10]$  (panel (d)). 

Panel (a) shows the diagram for $m\in(0,10]$ and fixed $F=0.7$. 
For $m=1.73$ a pitchfork bifurcation takes place, while for $m=1.78$ the inverse
pitchfork bifurcation occurs; both points are connected by two asymmetric stable
branches, while the main branch is unstable. We name this pair of pitchfork bifurcations
PB1. With a further increase of $m$ there are four more pairs of pitchfork and inverse
pitchfork bifurcations, for $m=2.14$, $m=2.44$ (PB2), $m=2.79$, $m=3.49$ (PB3),
$m=3.78$, $m=5.38$ (PB4) and $m=5.39$, $m=9.47$ (PB5). PB2 has the same structure as
PB1, hence the bifurcation points are connected by stable asymmetric branches. PB3 and
PB4 have a more complex structure. As an example we show PB3 in the zoom in
panel~(a1). The destabilization of the asymmetric solutions occurs through a sequence of period-doubling bifurcations leading to chaotic behavior. The last
asymmetric solution PB5 spreads over a much wider range of $m$; similarly, we observe a
period-doubling route to chaos and, additionally, destabilization of branches via
saddle-node bifurcations. 
The part of the branch present in the top-right corner of panel~(a) is only a small
part of another branch, because its right boundary is not in the considered range of $m$.
In summary, with the increase of $m$ the asymmetric branches
display more complex dynamics.

The two-parameter continuation in the $(F,m)$ space for $\omega=0.075$ and $\omega=1.0$ of the pitchfork bifurcations,
presented in panels~(b,c), confirms the results obtained in
Section~\ref{sec:Reduction-to-the}. To easily distinguish pairs of left and right bounds
of the bifurcations, we fill them with patterns and colors. As one can see, these
branches of pitchfork bifurcations converge, for $m\to0$, to the bifurcation point
$F_{bif}=\frac{2\sqrt{3}}{9}$. This point is the source of asymmetric solutions found here along
the symmetric periodic branches: when the left bound of PB1--5 crosses the symmetric
periodic solutions, a pitchfork bifurcation occurs and asymmetric solutions appear,
remaining until the crossing of the right bound. Within the
asymmetric ranges different types of solutions are present, starting from $1T$, then
$2T$ and its period-doubled successors $4T,8T,\dots$ (arising from period-doubling
bifurcations), together with odd multiples of $T$ (typically corresponding to windows in
the chaotic regime) which is visible in results from integration with Runge-Kutta method (not presented here). Additionally, we show the sixth pair of pitchfork bifurcations,
which starts for $F=0.7$ at $m=8.3$ and ends at $m=21.3$, hence we see only the bottom
part of the right bound. With increasing mass $m$ one additional branch appears.

 When forcing frequency is increased to $\omega=1.0$ the asymmetric solutions appear for smaller values of mass $m$, which is visible in scale of mass $m$ in panel~(c). To show what happens with further increase of $\omega$ we switched to $(m,\omega)$ space with fixed $F=0.7$ and performed continuation  of pitchfork tongues there. In zoom (panel (d1)) one can see that the tongues indeed occupy a progressively smaller range of $m$.

Because the system is invariant under $x\to-x$, $t\to t+\pi/\omega$, the
symmetric orbit treats both wells equivalently, while the asymmetric orbits
favor one well over the other. This well-favoring behavior can only be
sustained by inertia, so it vanishes as $m\to0$: every pitchfork tongue in
panels~(b,c) shrinks onto the single point $(0,F_{bif})$, which is therefore the
point from which all asymmetric solutions found here emerge.

\section{Conclusions}

We have shown that the transitions between the qualitatively different
regimes of the forced double-well Duffing oscillator have a simple
algebraic origin. In the massless limit the equation of motion
degenerates into a time-dependent cubic whose discriminant fixes the
number of instantaneous equilibria, and its vanishing defines the exact
threshold $F_{bif}$. This threshold leaves a measurable footprint at
finite mass: the intra-well/inter-well boundary converges to $F_{bif}$
as $m\to0$ and all obtained pitchfork bifurcation tongues generating asymmetric
inter-well solutions accumulate onto the single point $(0,F_{bif})$.
As $\omega$ increases, these tongues become more densely spaced,
accumulating over a progressively narrower interval of $m$. The
degenerate system thus anticipates the bifurcation structure of the
full oscillator and the massless limit acts as an organizing center of
Duffing dynamics.

The disappearance of the occupied branch relies only on a local extremum of the
static restoring force, which suggests that $F_{bif}$ and the convergence of the
intra-well/inter-well boundary to it may extend to other periodically forced
bistable systems. The accumulation of finite-mass pitchfork tongues onto
$(0,F_{bif})$, however, is established here only for the Duffing oscillator.

\section*{DATA AVAILABILITY}
The data that support the findings of this article are not
publicly available. The data are available from the authors
upon reasonable request.

\section*{Author Contributions}
 P.J. and P.P. conceived the study, developed the main ideas, interpreted the results, performed the numerical simulations and wrote the manuscript. P.J. developed and analyzed the massless-limit formulation. P.P. performed the bifurcation analysis and numerical continuation using AUTO-07p. Both authors approved the final manuscript.

\bibliographystyle{apsrev4-2}
\bibliography{bibtex}

\end{document}